\documentclass[a4paper]{jpconf}
\usepackage{graphicx}
\begin{document}
\title{Low angular momentum flow model of Sgr A* activity}

\author{B. Czerny, M. Mo\' scibrodzka} 

\address{Copernicus Astronomical Center, Bartycka 18, 00-716 Warsaw, Poland}

\ead{bcz@camk.edu.pl,mmosc@camk.edu.pl}

\begin{abstract}
Sgr A* is the closest massive black hole and can be observed with the highest angular resolution. Nevertheless, our current understanding of the accretion process in this source is very poor. The inflow is almost certainly of low radiative efficiency and it is accompanied by a strong outflow and the flow is strongly variable but the details of the dynamics are unknown. Even the amount of angular momentum in the flow is an open question. Here we argue that low angular momentum scenario is better suited to explain the flow variability. We present a new hybrid model which describes such a flow and consists of an outer spherically symmetric Bondi flow and an inner axially symmetric flow described through MHD simulations. The assumed angular momentum of the matter is low, i.e. the corresponding circularization radius in the equatorial plane of the flow is just above the innermost stable circular orbit in pseudo-Newtonian potential. We compare the radiation spectrum from such a flow to the broad band observational data for Sgr A*. 
\end{abstract}

\section{Introduction}
Accretion of matter onto a black is a complex non-stationary phenomenon, and an inflow of material is accompanied by an outflow. It is widely believed that magnetic fields play a key role but the details of accretion/outflow process are not understood.

Sgr A* is the closest massive black hole and therefore an obvious target for any future high spatial resolution observations. The Schwarzschild radius for the black hole mass of $3 \times 10^6 M_{\odot}$ at the distance of 8 kpc corresponds to $\sim 8 $ micro arc seconds. The next massive black hole (M87, distance of 16 Mpc, black hole mass $3 \times 10^9 M_{\odot}$) is by a factor of 2 smaller with respect to the angular size.

At the same time, Sgr A* is an example of a source with extremely low value of the Eddington ratio. It is by no means a spectacularly bright source, and the point like source emission in radio, IR and X-ray band was only detected and monitored in recent years. However, studying weakly active galaxies is important for two reasons: (i) galaxies spend most of their time in such phase (ii) the outflowing material may carry significant amount of kinetic energy so the feedback between the central activity and the galaxy itself may be also significant during the long inactive phase.
 
This studying Sgr A* is important but not easy. Models which apply to bright, high Eddington ratio sources like quasars may not necessarily apply to low level activity. We argue that accretion in Sgr A* is most likely an example of a low angular momentum flow, with circularization radius at a few or a few tenths of Schwarzschild radii, since such a flow is naturally variable and accompanied by a strong outflow so only a small fraction of the material actually reaches the horizon of a black hole.

\section{Accretion modes in Low Luminosity AGN}
Accretion pattern of the flow onto a black hole seems to be generally independent from the mass of the black hole but it is certainly sensitive to the dimensionless accretion rate, normalized with respect to black hole mass.
The spectra of high accretion rate sources like quasars or soft state galactic binaries are dominated by the optically thick geometrically thin Keplerian disk. Such a flow is radiatively very efficient, and the flow is well modeled by the standard Shakura \& Sunyaev (1973) disk.

The Keplerian disk forms due to two important aspects of the flow: (i) the angular momentum of the inflowing material is high (ii) the flow can cool efficiently. The first condition is somehow provided by the general flow setup. In case of binary systems, there is an angular momentum contained in the orbital motion of the two stars. The flow proceeds either through an inner Lagrange point (in low mass binaries) or through a highly focused wind (in high mass binaries, like Cyg X-1). In active galactic nuclei (AGN) the source of material is less specified but the similarity between the luminous AGN and galactic binaries suggests that also in this case the angular momentum of the material is high. It may come from a merger event, as major or minor mergers are usually considered as a reason for quasar activity. High angular momentum of the inflow helps to cool the flow since the angular momentum transfer is needed for accretion to proceed but nevertheless a significant local density is needed to provide considerable optical depth to the flow. The model is thus self-consistent for high accretion rates.

At the other extreme, we have Low Luminosity Active Galactic Nuclei (LLAGN), including Sgr A*, or galactic binaries in the quiescence state. It is highly unlikely that such sources contain cold radiatively efficient cold disks. In this case, the flow proceeds in radiatively inefficient way. Such flows can be grouped under the general name of RIAF - Radiatively Inefficient Accretion Flows.

The flow pattern in this case depends significantly on the outer boundary conditions, and the key parameter is the angular momentum of the material at the outer radius (see e.g. Czerny et al. 2007), which can be conveniently specified by giving the corresponding value of the circularization radius. From this point of view, the flows can be divided into three main categories:
\begin{itemize}
\item {spherical accretion}
\item {high angular momentum accretion}
\item {low angular momentum accretion.}
\end{itemize} 
Spherical accretion, or Bondi flow (Bondi 1952), is a specific case of inflow without any net angular momentum. The model applies if the circularization radius is smaller than the ISCO - Innermost Stable Circular Orbit. High angular momentum accretion, like advection-dominated accretion flow (ADAF; Ichimaru 1977, Narayan \& Yi 1994), advection-dominated inflow-outflow solutions (ADIOS; Blandford \& Begelman 1999) and similar specific models (e.g. convection dominated flows, Quataert \& Gruzinov 2000, disk-jet coupling models, Yuan et al. 2003) apply when the angular momentum at the outer radius is roughly Keplerian, $l_{out}/l_{K,out} \sim 1$ so the whole flow is strongly supported by the centrifugal force, and the efficient transport of angular momentum is needed at each radii for the accretion to proceed. The low angular momentum is an intermediate case where the outer part of the flow proceeds almost spherically while closer to the center a centrifugal barrier appears and the flow forms a kind of inner compact torus. In the case of a Schwarzschild black hole, the angular momentum at the ISCO is at
\begin{equation}
l_{ISCO} = {3 {\sqrt 3} \over {\sqrt 2} } {GM \over c},
\end{equation}
and an appropriate formula for a Kerr black hole can be found in Bardeen et al. (1972). Here $M$ is the black hole mass and $c$ the speed of light. The intermediate case:
\begin{equation}
l_{ISCO} < l < l_{K,out},
\end{equation}
is the most complicated one but its potential  applicability and consequences were also recognized quite early (Abramowicz \& \. Zurek 1981). 

\section{Analytical and numerical solutions for the dynamics of the flow in Sgr A*}

Analytical, or semi-analytical, models are convenient since they give a simple description of the flow pattern, they provide a good insight (since the underlying assumptions are controlled), and finally they have no restrictions to the model parameters within the applicability domain.
Numerical models can contain more physics, including the time-dependent evolution. On the other hand, they cannot extend arbitrarily both in time and space due to the practical limitations for the computing power. Therefore, the complementary use of both types of models is profitable and actually necessary. Both types of models were extensively used to explain the activity of Sgr A*.

\subsection{Analytical and semi-analytical models}

Spherically symmetric stationary Bondi model was one of the first detailed models of Sgr A* activity (e.g. Melia 1992, Melia et al. 2001), and it is still  frequently used to estimate the expected accretion rate from the temperature and the density measured roughly at the Bondi radius thanks to the high resolution of the Chandra satellite (e.g. Baganoff et al. 2003). However, the simple Bondi solution cannot extend neither too far nor too close to the black hole. At larger radii the role of an outflow is surely important. The amount of material available for accretion produced by the stellar winds ($\sim (10^{-6} -10^{-4}) M_{\odot}$yr$^{-1}$; Martins et al. 2007) is orders of magnitude higher than the amount of the inflowing material as recently measured from the Faraday rotation (Marrone et al. 2006). Thus, the outflow must be incorporated even into a spherical model, and the outflow dominates at distances larger than 2'' (Quataert 2004). At the smallest radii the requirement for no centrifugal force is likely to be unrealistic since the stars which are the source of the accreting material are in orbital motion (e.g. Rockefeller et al. 2004), although zero angular momentum flow from a single star is not excluded (e.g. Loeb 2004). The efficiency of the purely spherical flow also seems to be too low to explain the observed luminosity if the conservative estimate of the efficiency of direct electron heating is used (Mo\' scibrodzka 2006).

At the other extreme, there is a range of high angular momentum models, and some of them were used to interpret the broad band data in Sgr A*. An ADAF model was shown to be generally successful (e.g. Narayan, Yi \& Mahadevan 1995, Narayan et al. 1998) although it overproduced the amount of Faraday rotation. More advanced models taking into account the parametric description of the outflow (Yuan et al. 2003, 2004) corresponded better to the observational requirements (ADIOS parameter $s = 0.27$ used in this model suited best the data points).

Low angular momentum models were applied to Sgr A* by Mo\' scibrodzka et al. (2006). The model is based on the constant angular momentum flow models with possible shocks developed by Das 2002, Das et al. 2003, Chakrabarti \& Das 2004. The application of the model to Sgr A* gave an interesting result that, independently for a possibility of a shock existence, the stationary solution extended only to a certain radius (of order of ten - twenty $R_{Schw}$) and did not cover the whole flow. It clearly hinted that an inner non-stationary torus forms there and the accretion is a time-dependent process. This kind of model merges nicely with a separate class of models based on a priori assumption of the existence of an inner ring (Liu \& Melia 2002, Prescher \& Melia 2005, Tagger \& Melia 2006).  Such a ring is known to be unstable (Papaloizou \& Pringle 1984), and thus provides an attractive scenario for modeling Sgr A* outbursts. It is however interesting to note that while most papers concerning the inner torus treat its existence as an assumption, Mo\' scibrodzka et al. (2006) established its existence by analyzing the net angular momentum contained in stellar winds.  

\subsection{Numerical models}

Fully numerical, 2-D or 3-D, hydrodynamical (HD) or magnetohydrodynamical (MHD) models were also constructed and analyzed in the context of the flow in Sgr A*. However, due to limited possibility of the correct treatment of the outer and the inner part of the flow at the same time, the models addressed separately the issue of the outer or the inner part of the flow.

The outer part of the flow, including a realistic description of the stellar wind input to the circumnuclear material was first modeled Rockefeller et al (2004). It was a natural extension of the original work by Coker \& Melia (1997) who assumed randomly distributed 10 sources of material, without referencing to real stars at the vicinity of Sgr A* black hole. The model covered a distance range from the outer 3 pc to the inner $1.9 \times 10^{17}$ cm ($5 \times 10^5  r_g$). Inside this radius, the matter was expected to be in a free fall. This work was later continued e.g. by Cuadra et al. 2005, 2006, Coker \& Pittard 2007, with the most recent results obtained by Cuadra et al. 2008. This last work included the proper description of the motion of donor stars (including the possibility that stars form one or two inclined rings) as well as careful description of the gas cooling.

The second line of modeling focused on the inner flow description. These model do not describe the individual stellar input but adopt certain angular momentum distribution of the material at the outer edge of the computational domain and instead follow the inflow down to the black hole horizon. 

The zero angular momentum Bondi-Hoyle inflow was studied using the code {\sc zeus} by Coker \& Melia (1996). In their picture a highly supersonic wind flow from one direction arrives to the Sgr A* region and forms a bow shock. The accreted fraction of the material in this case has a net angular momentum corresponding to a circularization radius of $\sim 50 r_g$. This is cased by a highly non-uniform character of the flow. 
  
Significant fraction of the work concentrated on high angular momentum flows. However, in this case only the innermost part of the flow was modeled. 
The dynamics of the inner rotation-supported tori were studied numerically in the context of Sgr A* by Balbus \& Hawley (2002), Goldston et al. (2005) and Kato et al. (2005). These works contain, or were followed by certain estimates of the torus emissivity (e.g. Ohsuga et al. 2005 computations based on Kato et al. dynamics) but only local emissivity was included. Similar MHD computations with synchrotron emissivity were also performed by Chan et al. (2006). Mo\' scibrodzka et al. (2007) showed the resulting spectrum which included also non-local electron cooling; in this case the dynamics of the flow came from MHD simulations of Proga \& Begelman (2003). 

The restriction of the flow to the innermost region only results from the timescales present in such system. If the angular momentum is high the evolution at the outer edge of the computational zone is slow while the timestep must be short to resolve the inner flow.

\section{New hybrid MHD-Bondi model for Sgr A*}

\subsection{General set-up}

The exact MHD computations cannot cover large dynamical range in radii, so in the models well covering the innermost region the outer region of the flow modeled numerically is still well within the unresolved part of the Chandra image. Therefore, in order to make the comparison of the model with the data possibly accurate, we propose a new hybrid, analytical/numerical model of the flow.

We assume that at large distance the flow can be represented by a stationary spherically symmetric Bondi flow while in the inner part we perform 2-D MHD simulations in order to model the time dependent accretion flow.

The Bondi zone in our solution extends from the outer radius of $10^7 r_{Schw}$ to the transition radius fixed at 1200 $r_{Schw}$ (i.e. $1.3 \times 10^{19}$ cm and $1.6 \times 10^{15}$ cm correspondingly, for a black hole mass in Sgr A* of $3.6 \times 10^6 M_{\odot}$). The accretion rate in the Bondi flow is $7 \times 10^{-7}M_{\odot}$ yr$^{-1}$. The asymptotic sound speed at infinity is assumed to be $6.6 \times 10^7$ cm s$^{-1}$, and the asymptotic density is $2.2 \times 10^{-23}$ g cm${-3}$.

The conditions in the Bondi flow at the transition radius form boundary conditions for the inner MHD region. The radial velocity, the density and the energy density are assumed to be given by the outer Bondi solution. However, we allow for a non-zero angular momentum of the matter by assuming that the angular momentum density at the transition radius is equal
\begin{equation}
l = l_0(1 - cos \theta),
\end{equation} 
where $\theta$ is the angle measured from the symmetry axis and $l_0 = 1.36 l_{ISCO}$. The initial magnetic field is assumed to be vertical (i.e. parallel to the symmetry axis) and very weak, with the ratio of the gas to the magnetic pressure density of $10^{9}$ at the inner radius of the grid, and even higher further out. The inner radius of the MHD zone is set to 1.5 $r_{Schw}$. The dynamical timescale at the inner radius is 595 s.

 Inside this zone we perform MHD simulations with the code {\sc zeus}, as in Mo\' scibrodzka et al. (2007) and Mo\' scibrodzka \& Proga (2008). The simulations are performed in adiabatic approximation (i.e. radiative cooling is neglected), and the politropic index $\gamma = 5/3$ appropriate for a perfect gas is used (see Mo\' scibrodzka \& Proga 2008 for the discussion of the dependence of flow properties on this parameter). The pseudo-Newtonian potential is used to describe the gravitational field of the black hole.

\begin{figure}[h]
\begin{minipage}{16pc}
\includegraphics[width=16pc]{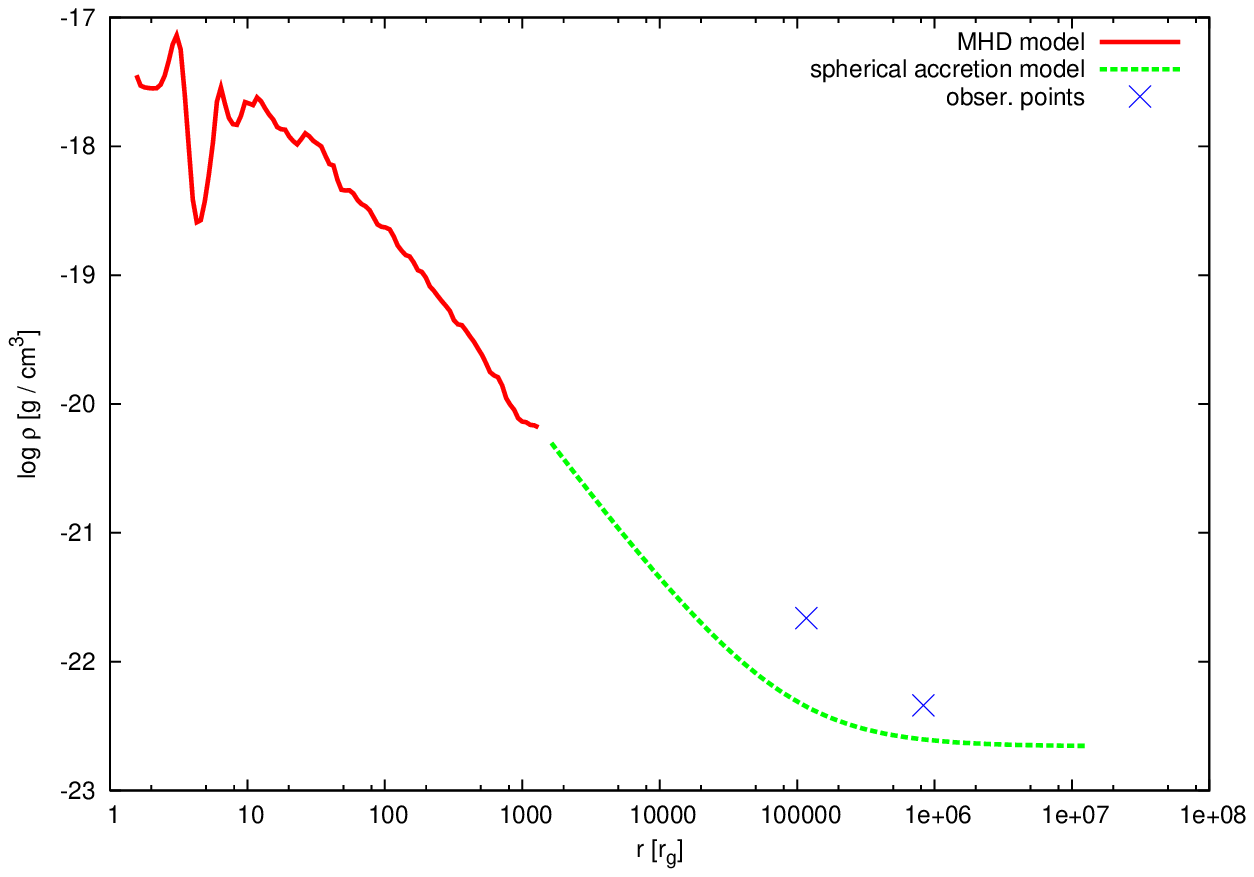}
\caption{\label{fig:profile_dens}The density profile for the combined numerical-MHD and analytical-Bondi solutions for the inflow in Sgr A* at the equatorial plane. Crosses mark the Chandra observational points from Baganoff et al. 2003.}
\end{minipage}\hspace{2pc}%
\begin{minipage}{16pc}
\includegraphics[width=16pc]{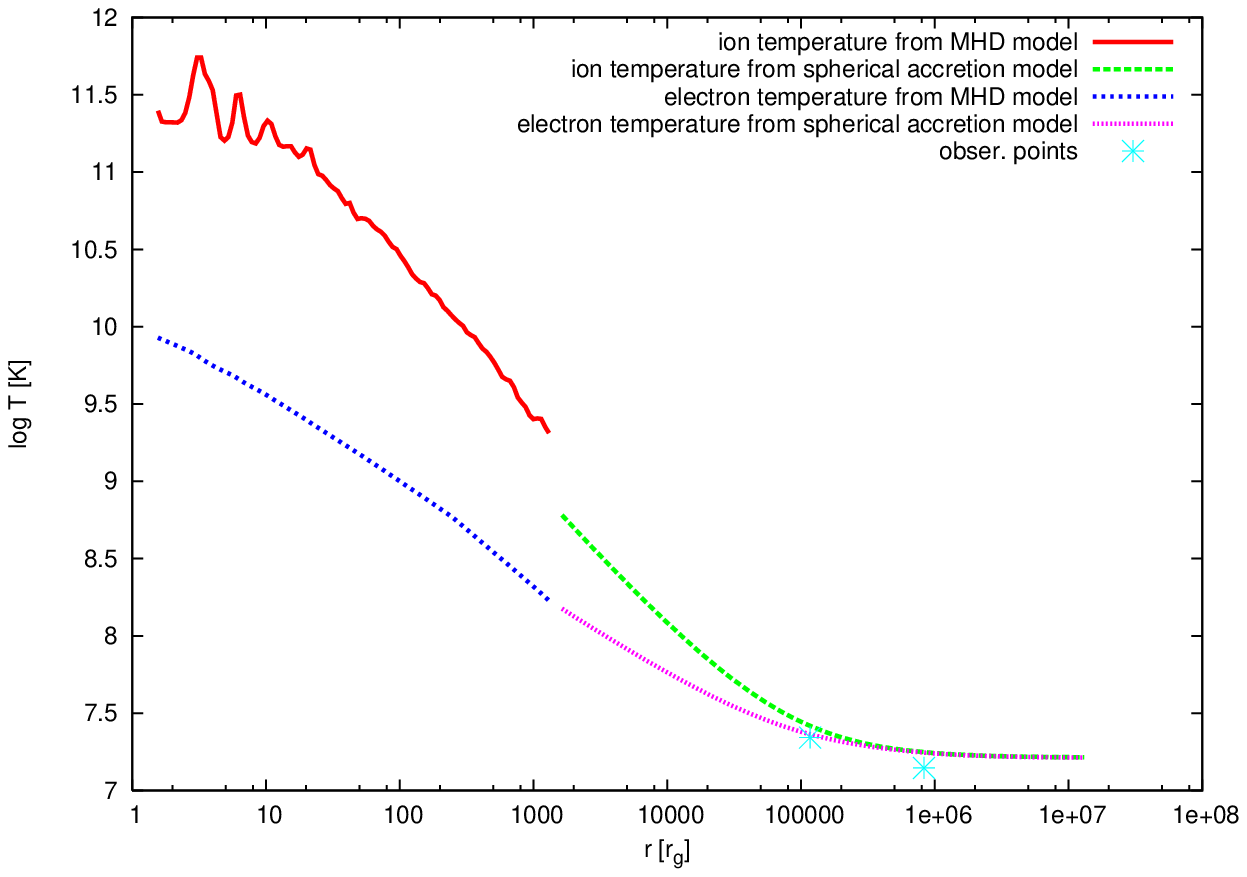}
\caption{\label{fig:profile_temp}The temperature ion and electron profiles for the combined numerical-MHD and analytical-Bondi solutions for the inflow in Sgr A* at the equatorial plane. Stars mark the Chandra observational points from Baganoff et al. 2003.}
\end{minipage} 
\end{figure}

\subsection{Results}

In Fig.~\ref{fig:profile_dens} and Fig.~\ref{fig:profile_temp} and we show the density and the ion temperature profile of the combined solution. The density profile matches quite smoothly across the transition region at the equatorial plane. The discrepancy becomes much larger closer to the symmetry axis since the pure flow outer Bondi solution does not take into account the outflow of material which develops in the inner MHD region calculated numerically. The outflow is highly non-uniform and time-dependent so the density contrast reaches at some locations/moments  of time up to two orders of magnitude. Of course it would be better to cover the whole range numerically but it is not realistic at present. 
Having the flow dynamics, we also calculate the representative broad band radiation spectrum from this simulation. We pick up one specific state of the flow corresponding to a fixed point in time, we treat it as a stationary solution and perform computations (i) of the electron temperature and (ii) radiative transfer.

The electron temperature distribution is obtained taking into account the electron-ion Coulomb coupling, the direct heating of the electrons due to gas compression and the advection of energy in electron flow. The direct electron heating is adopted to be $\delta = 0.01$. The advection is calculated using the electron temperature gradient at the transition radius from the Bondi flow. More details about the electron temperature computations are given in Mo\' scibrodzka et al. (2007). There is also slight discontinuity in the equatorial plane in the ion temperature distribution since this temperature in the inner region results from internal energy computation, and the energy is enhanced by strong turbulence. The electron temperature matches smoothly since the electron cooling is dominated by electron heat advection, and this advection is calculated a posteriori across the boundary.  
 
The radiation spectra are calculated using a Monte Carlo code developed by Mo\' scibrodzka (2006) and adopted to use with the numerical 2-D or 3-D flow as described in Mo\' scibrodzka et al. (2007).

The radiative transfer include synchrotron emission, bremsstrahlung and Comptonization. The magnetic field intensity in the inner region was taken self-consistently from the simulations, and in the outer Bondi region we assumed the gas pressure to magnetic field pressure ratio of 1 (i.e. equipartition condition). The thermal electron temperature was obtain self-consistently from the flow as described above but we additionally allowed for a small fraction of non-thermal electrons. They do not yet come from the model since they are accelerated in small unresolved shocks and/or magnetic field reconnections so their presence is described with the use of additional free parameters: a fraction of energy $\eta = 0.1$ is in the form of non-thermal electrons, and the maximum Doppler factor in the non-thermal electron distribution is $\gamma_{max} = 10^3$.

\begin{figure}[h]
\includegraphics[width=30pc]{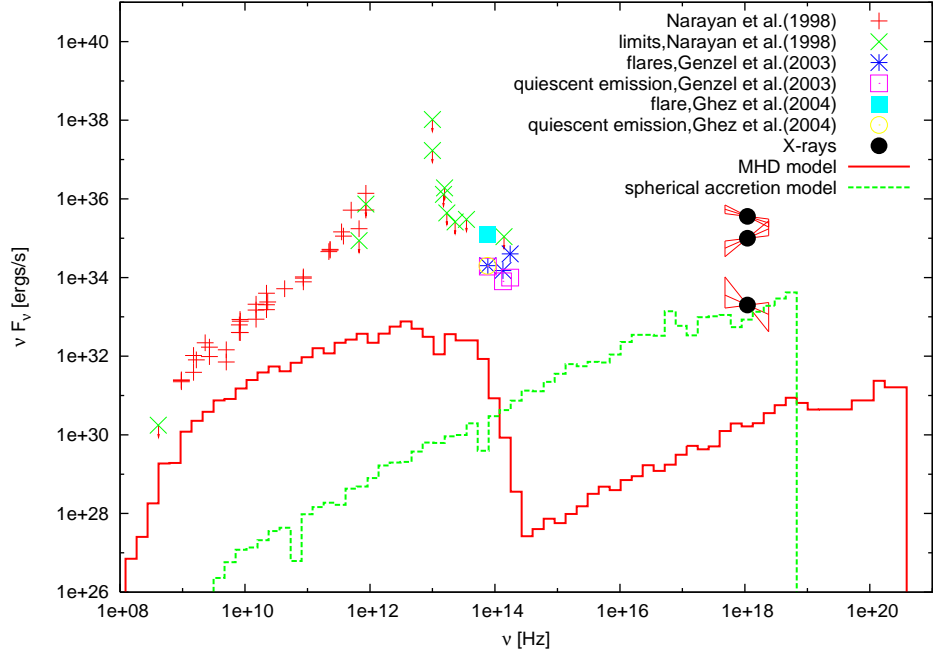}\hspace{2pc}%
\caption{\label{fig:spec}Exemplary broad band spectrum of Sgr A* obtained from a hybrid MHD-Bondi model representing a fixed moment of time in simulations.}
\end{figure}

The resulting broad band radiation spectrum, together with representative data points for Sgr A* is shown in Fig.~\ref{fig:spec}. The observational data are from Narayan et al. (1998), Genzel et al. (2003a), Ghez et al. (2004), Baganoff et al. (2003) and Porquet et al. (2003). The two contributions to the spectrum (inner MHD and outer Bondi) are shown independently to illustrate better where the emission originates. The total spectrum is the sum of the two contributions. The emission from the outer region is consistent with the level of the extended emission in the X-ray band. The nuclear emissivity in this model is too low to account well for the radio and mid-IR emission although the contribution of the non-thermal electrons is quite high. 

\begin{figure}[h]
\includegraphics[width=18pc]{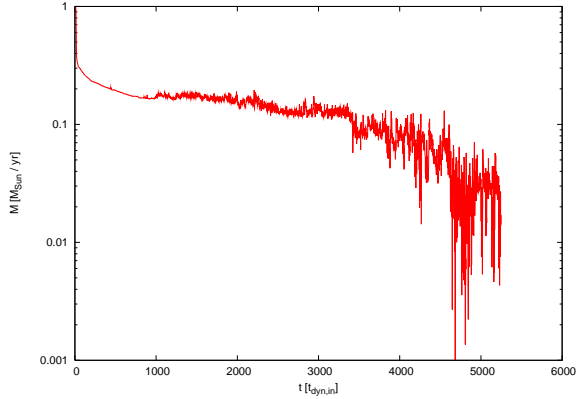}\hspace{2pc}%
\begin{minipage}[b]{16pc}\caption{\label{fig:accr} Accretion rate onto a black hole in Sgr A* as a function of time  band obtained from a hybrid MHD-Bondi model in units of $10^{-6}$.}
\end{minipage}
\end{figure}

The flow is generally variable so Fig.~\ref{fig:spec} is just an example of temporary spectrum. The accretion flow through the inner boundary of a grid varies in the simulations by up to two orders of magnitude, and the luminosity at various energy bands also changes. The effect is shown in Fig.~\ref{fig:accr}. The simulations were performed for over 5000 time bins in units of the dynamical timescale at the innermost edge of the computational zone, i.e. for 36 days in real time units. The flow already reached some form of equilibrium (see Mo\' scibrodzka \& Proga 2008 for computations performed for a slightly longer time). Plotting the spectrum we did not select a state in any specific way so the contribution from the inner MHD part of the flow can vary significantly. Nevertheless, it seems to fall systematically below the radio and mid-IR data points. 

\subsection{Discussion} 

This new hybrid MHD/Bondi model of the broad band spectrum is less successful in explanation of the observational data for Sgr A* than the model presented by Mo\' scibrodzka et al. (2007). The previous model fitted the data quite well. The difference is caused by higher accretion rate adopted in the previous model ($3.7 \times 10^{-6}M_{\odot}$ yr$^{-1}$) than in the current one ($7 \times 10^{-7}M_{\odot}$ yr$^{-1}$). The accretion rate ratio of a factor over 5 means almost a factor 30 difference in the luminosity since the local emissivity varies with the local density, $\rho$, as the $\rho^2$.  The previous model was not supplemented with the outer Bondi zone as it is done now; if we attempted to combine the previous model with such a spherically symmetric envelope the bremsstrahlung emission from the outer region over-predicted the level of the extended emission. The emissivity of a new model is nevertheless still somewhat lower than expected just at the basis of a simple scaling factor between the two values of the adopted accretion rates. It is due to the difference in the assumed distribution of the angular momentum as well as in the configuration of the initial magnetic field.
 
The spectra in Mo\' scibrodzka et al. (2007) were calculated for the flow dynamics computed as MHD simulations by Proga \& Begelman (2003). In those simulations the angular momentum distribution was described as constant from equatorial plane to $\theta = 45^{\circ}$, and the value of the constant was equal 2 $r_{Schw}c$, and it was zero at larger angles. In new simulation presented in this paper the value of the constant $l_0$ was 2.5 times higher but the angular momentum decreased with $\theta$ so effectively it was slightly higher.  The initial configuration of the magnetic field in Proga \& Begelman (2003) was purely radial while in the new simulation the initial magnetic field is vertical (parallel to the symmetry axis). The net average accretion rate in the model of Proga \& Begelman (2003) was by a factor of 0.025 smaller than a Bondi rate while in the new simulation the average inflow rate was smaller by a factor of 0.04 than the corresponding Bondi rate, but the Bondi rate itself was much smaller in the new MHD simulations as it was adjusted to the outer zone. More models should be calculated in order to disentangle the various effects.

The hybrid model itself is an interesting step towards the complete model covering both the inner region which determines the luminosity and the outer region which is marginally resolved in the observations. The presented solution is not yet successful since the outflow patters seen in the inner MHD solution should continue in the outer region. However, the solution in the inner region cannot be easily used as an inner boundary condition for MHD simulation of an outer region since they do not extend long enough in time. More complex, multi-timescale method should be applied to solve this problem. 

\section{Discussion}

The Sgr A* is certainly an extremely interesting object and a promising laboratory for the study of the accretion flow onto a black hole due to its unique large angular size. On the other hand, the accretion process in this source is particularly complex which is a characteristic property of the Low Luminosity AGN. The accretion/outflow pattern is highly non-stationary and it is likely to depend very strongly also on the outer/initial boundary conditions of the flow. This means that even with an advancement in the theory we cannot uniquely reconstruct the flow without direct observational information based on time-dependent imaging of the source with high spatial resolution. Thus Sgr A* is a much more difficult case to understand than bright quasars where a fairly standard Shakura-Sunyaev disk well accounts for most of the source properties.

This pessimistic aspect does not mean that there is nothing to be done in the meantime. As for the observations, it is important to continue the multi-wavelength monitoring as this gives direct constraints to the clumpiness of the flow as discussed by Baganoff and Eckart (this proceedings). It is also important to attempt to determine the density, temperature and emissivity map as close to the black hole as possible since the flow is not expected to be spherically symmetric at the Bondi radius. Instead, it is rather likely to show some axial symmetry, and the general level of the departure from a spherical symmetry is likely to measure the net angular momentum of the material in the circumnuclear region. On the theoretical side, simulations covering both the inner region and the outer resolved region are certainly needed, and the description of the outer region should contain the information about the motion of the donor stars as in Cuadra et al. (2008) as well as possible variations (time-dependence and clumpiness) of the stellar winds from individual stars. There are also many specific aspects of modeling which should be addressed. For example, the derivation of the electron temperature from the dynamical simulations in the inner MHD region require significant smoothing of the local energy density if the electron advection is also to be included (see Mo\' scibrodzka et al. 2007); otherwise, in some regions there is no self-consistent solution for the electron temperature. This means that some important physics is still missing in the models. 

So far the amount of angular momentum and its distribution in the accreting matter is an open question to be addressed in the future but in our opinion the models with low angular momentum content are most promising. In this case a kind of strongly variable inner torus forms in a natural way, and the strong short lasting outbursts are easily produced in such scenario. The torus itself, pushed by the continuously inflowing material, shows a kind of oscillations (Mo\' scibrodzka \& Proga 2008, in preparation) which may be of interest in the context of the quasi-periodic oscillations possibly seen in Sgr A* (in X-ray band: Genzel et al. 2003b, Aschenbach et al. 2004; in IR band: Meyer et al. 2006, Eckart et al. 2006, Trippe et al. 2007) although torus oscillations are of somewhat lower frequencies than $\sim 17$ minutes timescale seen in the data. 

\subsection*{Acknowledgments}
We thank Fred Baganoff and Daniel Proga for very helpful discussions. Part of this work was supported by grant 1P03D00829 of the Polish
State Committee for Scientific Research. M.M. acknowledges support provided by the Chandra award TM7-8008X issued by the Chandra X-Ray Observatory Center, which is operated by the Smithsonian Astrophysical Observatory for and on behalf of NASA under contract NAS 8-39073.

\smallskip

\end{document}